\documentclass[12pt]{iopart}

\expandafter\let\csname equation*\endcsname\relax
\expandafter\let\csname endequation*\endcsname\relax

\usepackage{graphicx}
\usepackage{amsmath}

\begin{document}

\title[Impact of Stimulated Raman Scattering on Dark Soliton Generation]{Impact of Stimulated Raman Scattering on Dark Soliton Generation in a Silica Microresonator}

\author{Gwangho Choi$^1$ and Judith Su$^{1,2}$}

\address{$^1$ Wyant College of Optical Sciences, The University of Arizona, Tucson, Arizona 85721, USA}
\address{$^2$ Department of Biomedical Engineering, The University of Arizona, Tucson, Arizona 85721, USA}
\ead{judy@optics.arizona.edu}

\begin{abstract}
Generating a coherent optical frequency comb at an arbitrary wavelength is important for fields such as spectroscopy and optical communications. Dark solitons which are coherent states of optical frequency combs in normal-dispersion microresonators can extend the operating wavelength and be excited via intermodal coupling. They have been investigated over the last decade due to their high conversion efficiency and deterministic excitation with no need for dispersion engineering. While the existence and dynamics of dark solitons has been examined extensively, requirements on the modal interaction for accessing the soliton state in the presence of a strong Raman interaction at near-IR wavelengths has been less explored. Here, analysis on the parametric and Raman gain in a silica microresonator is performed, revealing that parametric gain can be created by an additional frequency due to modal interaction and exceed the Raman gain. More complex interaction dynamics of the parametric and stimulated Raman scattering process is studied using numerical simulations based on the Lugiato-Lefever equation. It is found that exciting a dark soliton requires not only appropriate modal coupling but also a range of pump powers. The existence range of the dark soliton is analyzed as a function of pump power and detuning for given modal coupling conditions. We anticipate these results will benefit fields requiring optical frequency combs with high efficiency and selectable wavelength in a material with a strong Raman gain.
\end{abstract}

\vspace{2pc}


\maketitle

\section{Introduction} \label{sec:1}

The ultra-high quality (Q) factor and small mode volume of a microresonator greatly enhances the intracavity intensity in the microresonator and yields nonlinear effects such as stimulated Raman scattering (SRS) and four-wave mixing (FWM)~\cite{vahala_optical_2003,kippenberg_microresonator-based_2011,lin_review_2021}. While FWM is a parametric process where phase matching should be satisfied, SRS does not require phase matching~\cite{spillane_ultralow-threshold_2002,kippenberg_kerr-nonlinearity_2004}. Engineering the  dispersion of the cavity and choosing proper experimental parameters can excite FWM over SRS, and generate optical frequency combs~\cite{kippenberg_kerr-nonlinearity_2004, delhaye_optical_2007, agha_four-wave-mixing_2007, savchenkov_tunable_2008}. The FWM process can initiate a Kerr frequency comb and lead to soliton generation in microresonators with a proper choice of power and detuning~\cite{herr_universal_2012,herr_temporal_2014,guo_universal_2017}. 
A  bright soliton which is a coherent state of an optical Kerr frequency comb in the anomalous dispersion regime can be soft-excited inherently (i.e., the soliton state can be reached with a continuous wave (cw) background)~\cite{hansson_dynamics_2013,herr_temporal_2014}. In contrast, dark solitons may be soft-excited via intermodal interaction~\cite{xue_mode-locked_2015,fulop_high-order_2018,nazemosadat_switching_2021} or aid of an auxiliary resonator~\cite{miller_tunable_2015,kim_dispersion_2017,kim_turn-key_2019, helgason_dissipative_2021}, and hard-excited (i.e., the soliton state cannot be reached with the cw background and may require manipulation of the background) by a modulated pump~\cite{liu_bright_2017, lobanov_generation_2019, liu_stimulated_2022} or self-injection locking~\cite{liang_generation_2014, wang_self-regulating_2021,lihachev_platicon_2021} in the normal dispersion regime. 

While optical microresonators can be designed to possess anomalous dispersion at near-IR wavelengths by engineering the geometry of the resonator, often this requires precise fabrication control or additional fabrication processes (e.g., incorporation of a particular coating)~\cite{fujii_dispersion_2020}. Anomalous dispersion, however, can also be created locally via interaction of different optical mode families supported in the resonator. This can occur regardless of the dispersion of the cavity and operating wavelength~\cite{liu_investigation_2014,xue_mode-locked_2015}. Since WGM resonators such as microtoroids and spheres can support a greater number of optical modes compared to integrated microring resonators,they can introduce modal interaction without precise fabrication techniques.
Thus, in this paper, we only focus on the mode-interaction-aided excitation method which may be readily implemented on WGM resonators (e.g., microtoroid or microsphere resonators) that are an attractive platform due to their higher Q factor and do not need ultra-fine fabrication techniques as their surface roughness can be greatly reduced by a thermal reflow process~\cite{braginsky_quality-factor_1989,armani_ultra-high-q_2003}. Note that a higher Q factor not only decreases the threshold power for nonlinear effects but is beneficial in applications, such as biosensing~\cite{su_label-free_2016,choi_optical_2022-1, ozgur_ultrasensitive_2019-1, li_dark_2019, chen_simulating_2019, suebka_how_2021, hao_noise-induced_2020, dellolio_photonic_2021, lu_split_2013}. 

SRS can lead to Raman lasing by pumping a resonance above its SRS threshold power regardless of the dispersion of the cavity~\cite{spillane_ultralow-threshold_2002, kippenberg_ultralow-threshold_2004}. 
Although engineering dispersion of a cavity can make the FWM process dominant over the SRS process in the anomalous dispersion regime, there may still be effects of the Raman interaction including Raman self-frequency shift~\cite{karpov_raman_2016} and Stokes solitons~\cite{yang_stokes_2017, tan_multispecies_2021}. In crystalline materials where the Raman gain has a narrow bandwidth, SRS can be avoided by not overlapping the Raman gain and a mode of a cavity~\cite{okawachi_competition_2017,xia_engineered_2022}. The interaction between FWM and SRS can also yield effects such as Raman combs~\cite{lin_phase-locking_2016,suzuki_broadband_2018}, broader Kerr frequency combs~\cite{weng_octave-spanning_2021,liu_integrated_2018}, and Raman-assisted FWM~\cite{yao_generation_2020}. 

The transition and competition between SRS and FWM has been studied in the context of frequency detuning between a pump laser frequency and a resonant frequency, coupling conditions, and geometrical factors ~\cite{kippenberg_kerr-nonlinearity_2004,min_controlled_2005,agha_four-wave-mixing_2007,grudinin_impact_2013,yang_transition_2021}. The transition from Raman oscillation to FWM based parametric oscillation was reported in these works, but their analysis is limited to comparing the gains (or threshold powers) for both phenomena. In fact, complex dynamics of these nonlinear effects can be better understood by considering their interactions combined with discrete resonance modes separated by a free-spectral range (FSR) in a microcavity~\cite{chembo_spatiotemporal_2015,torres-company_comparative_2014,fujii_transition_2018,yang_stability_2022}. While there are a number of studies on this interaction in optical resonator systems in the anomalous dispersion regime~\cite{milian_solitons_2015, lin_phase-locking_2016, lin_universal_2016, karpov_raman_2016, bao_observation_2016, yang_stokes_2017, kato_transverse_2017, liu_integrated_2018, wang_stimulated_2018,  yao_generation_2020, gong_near-octave_2020, weng_octave-spanning_2021}, only a limited number of studies focus on this in the normal dispersion regime~\cite{cherenkov_raman-kerr_2017, yao_generation_2020, liu_dynamics_2021, parra-rivas_influence_2021}. This is partly due to its difficult excitation in experiments~\cite{hansson_dynamics_2013, xue_mode-locked_2015}.
Although the excitation dynamics of dark solitons~\cite{jang_dynamics_2016, fujii_analysis_2018, nazemosadat_switching_2021} and the influence of SRS on dark solitons~\cite{parra-rivas_influence_2021} has been investigated, the complex interaction of SRS and dark solitons and their excitation dynamics has been less explored. Furthermore, in a material with a strong Raman gain, dark soliton generation may be significantly perturbed by SRS. This will, in turn, yield more limited conditions for both the excitation and stability region of the dark soliton.

In this work, we numerically study the excitation or accessibility of dark solitons in the presence of Raman interactions in a normally dispersive microresonator at near-IR (NIR) wavelengths~(here, $780~\mathrm{nm}$). Since the Raman gain ($g_\mathrm{R}$) at this wavelength is twice as big as at IR wavelengths~(i.e., $g_\mathrm{R}~(\lambda\!=\!0.78\!~\mathrm{\mu m})\approx2g_\mathrm{R}~(\lambda\!=\!1.55\!~\mathrm{\mu m})$), the interaction may be even more complex~\cite{newbury_pump-wavelength_2003,lin_raman_2006}. It was found that an additional frequency shift caused by an avoided-mode-crossing (AMX) can create parametric gain whose amplitude and bandwidth are dependent on both location and amplitude of the AMX. In case the pump power is below the threshold power for SRS, FWM can be initiated and a dark soliton can be generated with a proper AMX condition. Moreover, even if the pump power is above the threshold power for SRS, a dark soliton can still exist but under more restricted conditions. We first study how parametric gain can be introduced by the mode-interaction (or AMX) and compare the parametric gain with the Raman gain with different simulation parameters in section~\ref{sec:2}. Next, we numerically simulate a dark soliton under fixed parameters~(section~\ref{sec:3}). In section~\ref{sec:4}, we discuss in detail interactions of FWM and SRS under different conditions. Finally, stimulated stability charts are presented in section~\ref{sec:5}. 
\section{Gain curves for FWM based parametric oscillation and stimulated Raman oscillation} \label{sec:2}

Raman gain exists regardless of the dispersion of a cavity, and can stimulate Raman oscillation with no phase matching condition satisfied if it is externally pumped beyond its threshold power~\cite{kippenberg_kerr-nonlinearity_2004}. The Raman gain per roundtrip, $g_{\mathrm{R}}$, in silica can be expressed as follows~\cite{newbury_pump-wavelength_2003,agrawal_nonlinear_2019, fujii_transition_2018}:
\begin{equation} \label{eq:1}
g_{\mathrm{R}} = \alpha + g_{\mathrm{bulk}}^{R}\frac{P_0}{A_{\mathrm{eff}}}L_{\mathrm{eff}},
\end{equation}
where $\alpha$ is the roundtrip loss, $g_{\mathrm{bulk}}^{R}\approx1.3\times10^{-13}~\mathrm{m/W}$ is the bulk Raman gain of silica at $780~\mathrm{nm}$, $A_\mathrm{eff}$ is the effective mode area, $L_{\mathrm{eff}}=(\alpha/L)^{-1}(1-\exp(-\alpha))$ is the effective length, $L$ is the length of the cavity, and $P_0$ is the intracavity power which can be obtained by the following equation~\cite{agrawal_nonlinear_2019,fujii_transition_2018}:
\begin{equation} \label{eq:2}
(\gamma L)^2P_0^3 - 2\delta_0\gamma LP_0^2 + (\delta_0^2+\alpha^2)P_0 = \theta P_{\mathrm{in}},
\end{equation}
where $\delta_0=t_R(\omega_0-\omega_\mathrm{p})$ is the phase detuning of the pump frequency ($\omega_\mathrm{p}$) with respect to the nearest resonant frequency ($\omega_0$), $t_R$ is the cavity roundtrip time, $\gamma=n_2\omega_0/(cA_\mathrm{eff})\approx 0.014$ is the nonlinear coefficient, $n_2$ is the nonlinear refractive index, $c$ is the speed of light in vacuum, $\theta$ is the coupling coefficient between the cavity and waveguide, and $P_{\mathrm{in}}$ is the pump power. Note that the Raman gain is linearly dependent on the intracavity power which can be determined by choosing a detuning and pump power for a cavity.

In the anomalous dispersion regime, the parametric gain of a cavity, $g_\mathrm{cav}$, can be created and expressed by the equation~\cite{kippenberg_kerr-nonlinearity_2004, torres-company_comparative_2014, fujii_transition_2018}
\begin{equation} \label{eq:3}
g_{\mathrm{cav}}(\Omega) = \alpha + \sqrt{(\gamma LP_0)^2-(\delta_\mathrm{mis})^2},
\end{equation}
where $\delta_\mathrm{mis}=\delta_0-(\beta_2/2)L\Omega^2-2\gamma LP_0$ is the phase-mismatch due to the detuning, dispersion, and nonlinearity, $\beta_2$ is the second-order dispersion coefficient ($\beta_2<0$ in  anomalous dispersion regime), $\Omega=bD_1$ is the modulation frequency, $b$ is the mode number with the additional frequency shift~($b=1$, unless otherwise stated, for simplicity), and $D_1=2\pi\mathrm{FSR}$ is the FSR in angular frequency at $\omega_0$. 

\begin{figure}[tb!]
    \centering
    \includegraphics[width=\textwidth]{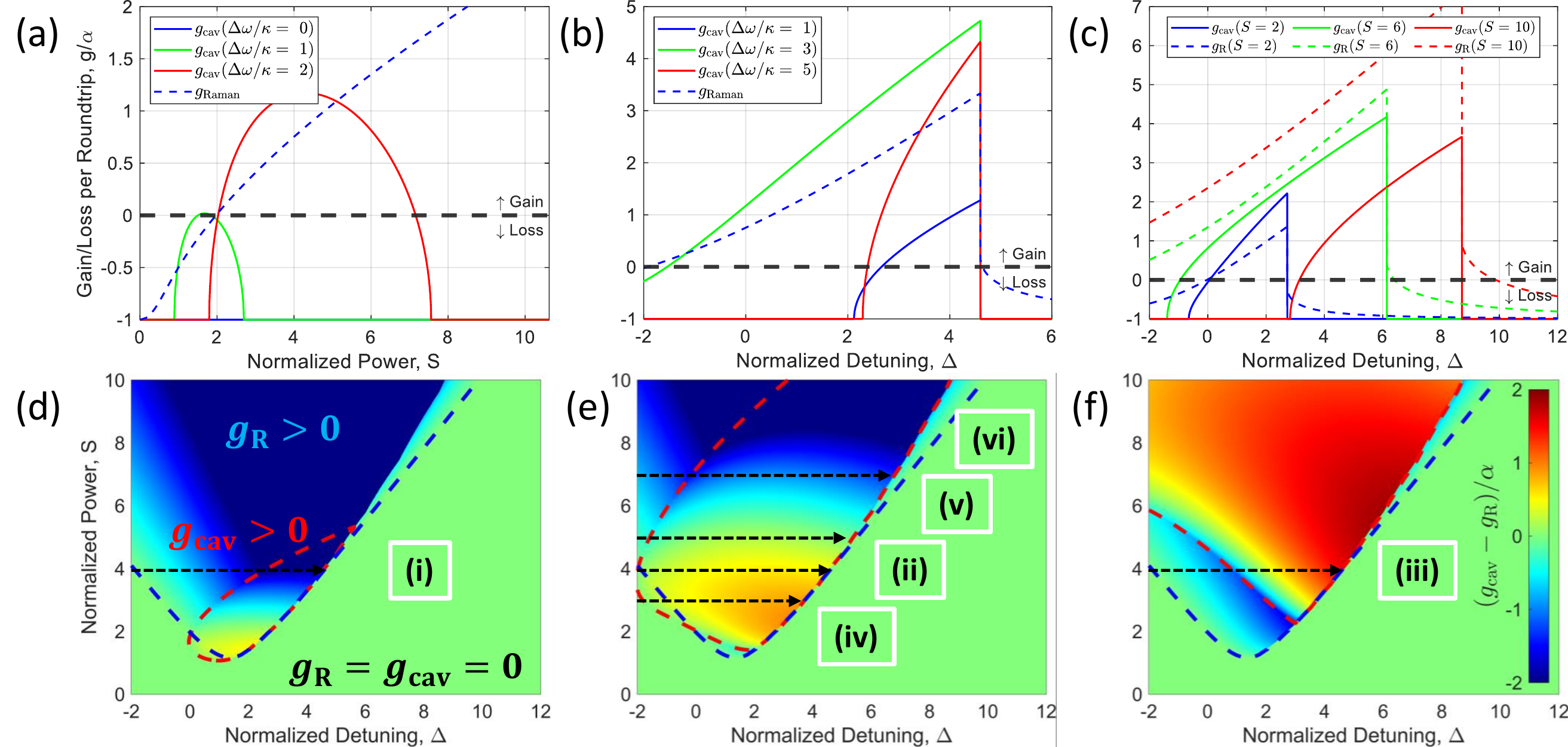}
    \caption{Parametric four-wave mixing (FWM) and stimulated Raman scattering (SRS) gain curves. (a - c) Gain/loss per roundtrip vs (a) normalized power and (b), (c) normalized detuning in the normal-dispersion regime. (a) Detuning ($\delta_0$) is fixed at 0. No gain is present in the absence of modal interaction. Parametric gain can be created by introducing modal interaction~($\Delta\omega>0$), which determines an amplitude and width of the gain envelope, and a threshold power. (b) Normalized power ($S$) is fixed at 4. Raman gain is not dependent on the frequency shift. Parametric gain is maximized at $a\approx3$. (c) Parametric and Raman gains at different pump powers with a fixed frequency shift ($a\approx2$). Raman gain increases linearly with the pump power, while parametric gain can be a function of both the pump power and additional frequency shift. (d - f) Difference between the FWM and SRS gains in 2D-parameter space. The FWM (SRS) dominant region is filled with red (blue). Red (blue) dashed line represents zero gain for FWM (SRS). Horizontal dashed arrows indicate excitation pathways explored in upcoming sections. The difference between the FWM and SRS gains is normalized by loss ($\alpha$) with the chosen additional frequency shifts of (d) $a=1$, (e) $a=2$, and (f) $a=4$. Cases~(i - vi) shows parameters analyzed in the following sections. Note $b$ is assumed to be 1 in all calculations.}
    \label{fig:1}
\end{figure}

The parametric gain created by AMX may be expressed by adding an additional phase shift in the normal dispersion regime ($\beta_2>0$). The phase-mismatch term then becomes
\begin{equation} \label{eq:4}
\delta_\mathrm{mis}=\delta_0-(\beta_2/2)L\Omega^2-2\gamma LP_0 + \Delta\delta
\end{equation}
where $\Delta\delta=\Delta\omega t_R$ is the additional phase shift per roundtrip, $\Delta\omega=a\kappa$ is corresponding angular resonance frequency shift, $a$ is the normalization factor of the additional frequency shift, and $\kappa/2\pi$ is the FWHM of the resonance of the cavity. The second term in equation~(\ref{eq:4}) which is negative in normal dispersion regime may be compensated by the additional frequency shift due to AMX~(i.e.,~$\Delta\delta>0$). Figure~\ref{fig:1}(a) shows both the normalized parametric gain and Raman gain at different additional frequency shifts as a function of normalized power, $S=\sqrt{\gamma L \theta P_\mathrm{in} / \alpha^3}$, at a fixed detuning, $\delta_0=0$. Raman gain is not dependent on the additional frequency shift and it remains the same. Interestingly, parametric gain can be created by the $\Delta\omega$. It was found that the threshold power, existence range, and maximum gain for the FWM process are dependent on the $\Delta\omega$. The threshold power tends to increase linearly with the addition frequency shift, while the existence range and maximum gain hit a maximum at a certain $\Delta\omega$ value.

In practice, only the detuning is swept from high to low frequency instead of the pump power to access a 'thermal triangle'~\cite{carmon_dynamical_2004}. Thus, it may be straightforward to plot the gain curves as a function of detuning from the blue to red-detuned side. Figure~\ref{fig:1}(b) shows the same gain curves as a function of the normalized detuning, $\Delta=\delta/\alpha$, at a fixed normalized power, $S=4$, for different $\Delta\omega$ values. The parametric gain is created at a small additional frequency shift~($a=1$), maximized at a certain point~($a=3$), and shrinks at a large frequency shift~($a=5$). It may be found that a certain amount of  additional frequency shift is required to overcome the loss in the cavity, i.e., $g_\mathrm{cav}>0$. In addition, there can be FWM dominant regions over SRS for certain frequency shifts, i.e., $g_\mathrm{cav}>g_\mathrm{R}>0$. Figure~\ref{fig:1}(c) presents the same gain curves as a function of detuning at a fixed frequency shift ($\Delta\omega=2\kappa$) for different normalized powers. The Raman gain curves increase with pump power, while the parametric gain is bigger at $S=6$ than other cases. SRS dominates over FWM at relatively high pump powers; however, under a proper frequency shift condition it is possible that FWM can overcome SRS at relatively low pump power. A direct comparison between parametric and Raman gain is shown in two-dimensional parameter space at different additional frequency shifts in Figures~\ref{fig:1}(d - f). The red (blue)-colored region represents the larger parametric (Raman) gain region. The Raman gain (blue dashed region) remains the same, while the parametric gain (red dashed region) region gets bigger as a function of the additional frequency shift, but shrinks after a maximum point.
\section{Numerical model} \label{sec:3}

The intracavity field of the microsresonator can be modeled by the well-known Lugiato-Lefever equation (LLE) as follows~\cite{chembo_spatiotemporal_2013,chembo_spatiotemporal_2015,parra-rivas_influence_2021}:

\begin{multline} \label{eq:5}
t_R\frac{\partial E}{\partial t} = -(\alpha  + i\delta_0) E + \sqrt{\theta} E_{in} - i\frac{\beta_2 L}{2}\frac{\partial^2}{\partial\tau^2}E\\
+ i\gamma L(1-f_\mathrm{R})|E|^2E + i\gamma Lf_\mathrm{R}(R \ast |E|^2)E
\end{multline}
where $E(t,\tau)$ is the internal electric field within the resonator, $t$ is the slow time describing the evolution of the field envelope, $\tau=t_R(\phi/\pi)$ is the fast time describing the temporal profile of the field envelope, and $\phi$ is the azimuthal coordinate around the resonator. $f_\mathrm{R}$ is the fractional coefficient which determines the strength of the SRS term, and $*$ denotes the convolution. $f_\mathrm{R}$ is assumed $0.18$ for silica~\cite{lin_raman_2006}. $R(\tau)$ is the Raman response function
\begin{equation} \label{eq:6}
R(\tau)=\frac{\tau_1^2+\tau_2^2}{\tau_1\tau_2^2}\exp^{-\tau/\tau_2}\sin{(\tau/\tau_1)}
\end{equation}
where $\tau_1=12.2~\mathrm{fs}$ and $\tau_2=32~\mathrm{fs}$ for fused-silica based fibers~\cite{lin_raman_2006}. A complex dispersion profile of a microresonator without AMX can be described in the frequency domain as follows: $D_\mathrm{int}=\omega_\mu-(\omega_0+D_1\mu)=\frac{1}{2}D_2\mu^2+\ldots$, where $D_\mathrm{int}$ is the integrated dispersion, and $\omega_\mu$ is the angular frequency of the relative mode number~($\mu$) with respect to the pump mode~($\mu=0$). Note that we ignore higher-order ($\beta_{i>2}$ or $D_{i>2}$) dispersion coefficients to simplify simulations and focus on effects of AMX and SRS. The integrated dispersion with the AMX effect may be simply expressed as~\cite{herr_mode_2014}
\begin{equation} \label{eq:7}
D_\mathrm{int}(\mu,a,b)=\omega_\mu-(\omega_0+D_1\mu)=\frac{1}{2}D_2\mu^2-\frac{a\kappa/2}{\mu-b-0.5}
\end{equation}
where $a$ and $b$ determine the normalized amplitude and the location of the additional frequency shift. Note this model describes the dispersion for resonators with a strong intermodal coupling, while adding a single additional frequency shift for a specific mode number better describes resonators with a weak AMX~\cite{xue_mode-locked_2015,bao_observation_2018}. Then the LLE may be rewritten by taking the Fourier transform and the inverse Fourier transform of the dispersion and Raman terms:
\begin{multline} \label{eq:8}
t_R\frac{\partial E}{\partial t} = -(\alpha  + i\delta_0) E + \sqrt{\theta} E_{in} - i\mathcal{F}^{-1}\left[t_RD_\mathrm{int}\cdot\mathcal{F}[E]\right]\\
+ i\gamma L(1-f_\mathrm{R})|E|^2E + i\gamma Lf_\mathrm{R}(\mathcal{F}^{-1}\left[\mathcal{F}[R]\cdot\mathcal{F}[|E|^2]\right])E
\end{multline}
where $\mathcal{F}$ and $\mathcal{F}^{-1}$ denote the Fourier transform and the inverse Fourier transform, respectively. The LLE is solved numerically using the split-step method where the nonlinear and dispersion contributions are treated separately~\cite{agrawal_nonlinear_2019}. 

\begin{figure}[tb!]
    \centering
    \includegraphics[width=\textwidth]{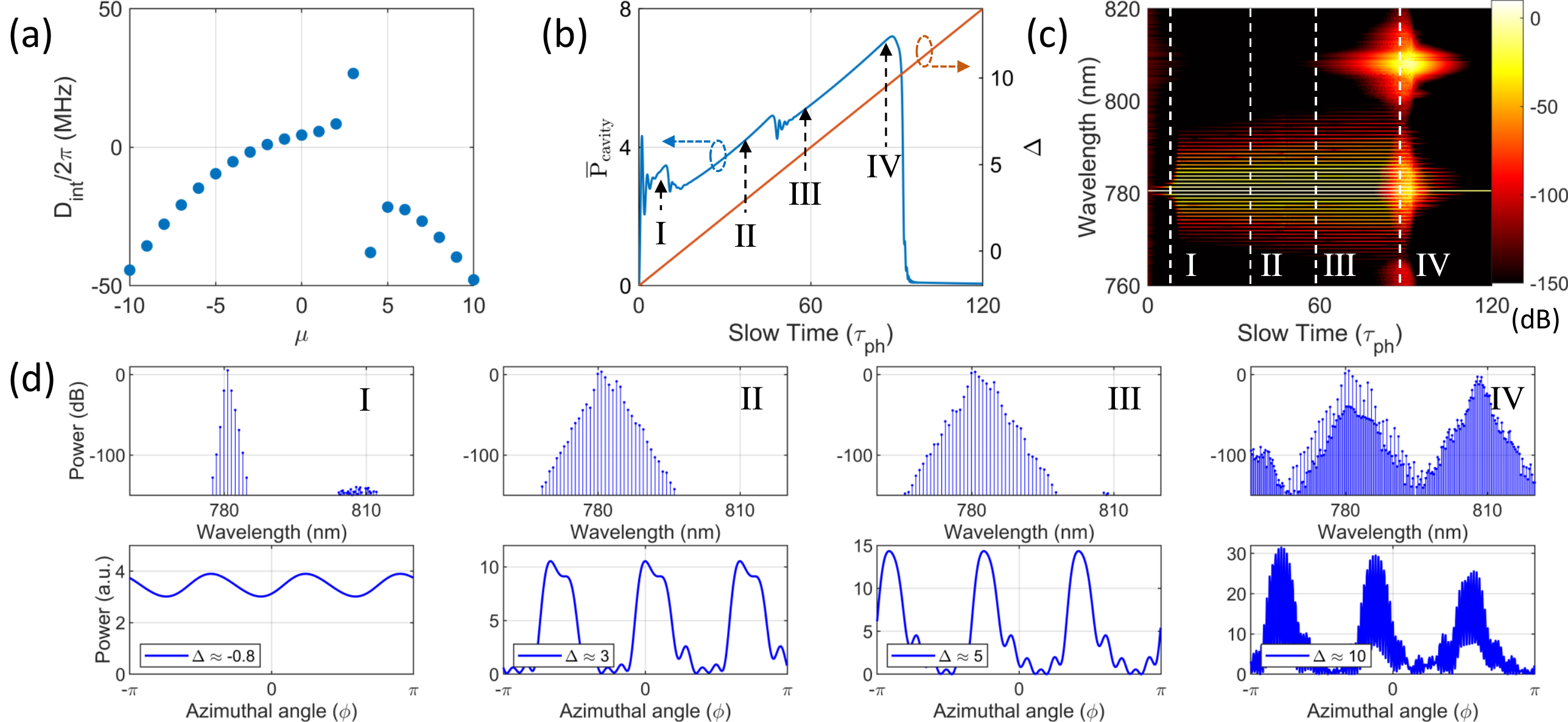}
    \caption{Excitation of dark soliton and SRS. (a)~The integrated dispersion with an AMX~($a=8,b=3$) based on equation~(\ref{eq:7}). (b)~The averaged intracavity power (blue) and detuning~(orange) as a function of time. The normalized pump power set to 4. (c) The spectral evolution of the intracavity power. SRS is excited when the intracavity power reaches the threshold intracavity power. (d) The spectrum and temporal profile at the stages marked in (b). Four stages are chosen at different detuning values.}
    \label{fig:2}
\end{figure}

We consider a silica microtoroid resonator with a radius of $250~\mathrm{\mu m}$ at $780~\mathrm{nm}$ for LLE simulations. The simulation parameters are set as follows: $D_1/2\pi=130.4~\mathrm{GHz}$, $D_2=-5.72~\mathrm{MHz}$, $Q_\mathrm{load}=1\times10^8$, $\alpha=t_R\omega_0/2Q_\mathrm{load}=9.25\times10^{-5}$, $\theta=2.71\times10^{-5}$, $\gamma=0.014$, and $S=4$. The integrated dispersion with a AMX, $D_\mathrm{int}(\mu,a=8,b=3)$, is shown in figure~\ref{fig:2}(a). The normalized detuning is linearly increased over time from -2 to 14 to scan the resonance from the blue-detuned side to the red-detuned side which is usually done in real experimental situations. The corresponding average intracavity power is shown in Figure~\ref{fig:2}(b). The spectral evolution of the intracavity field is shown in Figure~\ref{fig:2}(c). The spectral and temporal profile are plotted in Figure~\ref{fig:2}(d) at different detuning values which are indicated as vertical dashed lines in Figures~\ref{fig:2}(b) and (c). Unlike its counterpart bright soliton where 'step-like' patterns indicate transition to soliton states in the effectively red detuned side~\cite{herr_temporal_2014,yi_soliton_2015}, dark solitons can be accessed in the effectively blue detuned side~\cite{xue_mode-locked_2015,nazemosadat_switching_2021,helgason_dissipative_2021}. Dark soliton states can be determined by their temporal profiles which indicate pulse-like patterns. 

As discussed in section~\ref{sec:2}, the AMX effect may generate the parametric gain. Here we focus on the excitation pathway corresponding to the case ($\mathrm{\textbf{ii}}$) as labeled in Figure~\ref{fig:1}(e). In this case, it is expected that the FWM process is dominant over the SRS process because the parametric gain is bigger than the Raman gain. But as $\Delta$ increases, the intracavity power also increases and generates strong Raman gain along with the parametric gain. Thus, some complex interaction or competition between them may be expected. At stage $\mathrm{I}$ in Figure~\ref{fig:2}(d), it is shown that FWM comb can be initiated and leads to a Turing pattern~\cite{xue_mode-locked_2015,nazemosadat_switching_2021}. The first sideband location coincides with the AMX location~(here, $\mu=b=3$)~\cite{liu_investigation_2014}. As the pump wavelength increases, the bandwidth of the comb increases and a 'step-like' pattern in the blue-detuned side is observed indicating a transition to a coherent state as reported in~\cite{xue_mode-locked_2015,nazemosadat_switching_2021, xia_soliton_2022}. Localized structures in the cavity are observed as the detuning is increased~(stages $\mathrm{II}$ and $\mathrm{III}$ in figure~\ref{fig:2}(d)). The number of localized structures is equivalent to the AMX location. We also observed that the number of low intensity oscillations at the dark pulse profiles increases at a function of the detuning (i.e., 4 and 5 oscillations at $\Delta=3$ and $5$, respectively) as it is predicted theoretically~\cite{parra-rivas_dark_2016, parra-rivas_origin_2016} and verified experimentally~\cite{xue_mode-locked_2015,nazemosadat_switching_2021}. At a large detuning, the intracavity power is high enough to initiate SRS and the Raman oscillation gets dominant~(stage $\mathrm{IV}$ in figure~\ref{fig:2}(d)). Note that the SRS gets dominant at a lower intracavity power for a large $f_\mathrm{R}$.
\section{Results and discussions} \label{sec:4}

\subsection{Influence of AMX on dynamics of dark soliton generation} 

While suppressing the AMX may simplify and help the excitation of bright solitons in anomalous-dispersion microresonators~\cite{herr_mode_2014}, AMX is required to soft-excite a FWM comb~\cite{savchenkov_kerr_2012} and may lead to dark soliton states in normal-dispersion microresonators~\cite{liu_investigation_2014,xue_mode-locked_2015}. We study three cases (corresponding to cases ($\mathrm{\textbf{i}}$), ($\mathrm{\textbf{ii}}$), and ($\mathrm{\textbf{iii}}$) as labeled in Figures~\ref{fig:1}(d), (e), and (f), respectively) where different excitation dynamics of both the dark soliton and SRS are expected in each case. Again the Raman gain is not dependent on the AMX, while the amplitude and bandwidth of the parametric gain are dependent on the magnitude and location of the AMX as shown in Figure~\ref{fig:1}(b). The first case~($\mathrm{\textbf{i}}$) shows the parametric gain is not enough to overcome the Raman gain and the Raman oscillation is a dominant effect. The second case~($\mathrm{\textbf{ii}}$) is where the parametric gain is bigger than the Raman gain and the FWM is the dominant process, but there may be a gain competition as the intracavity power grows. More complex dynamics is observed as in case~($\mathrm{\textbf{iii}}$) where the parametric gain envelope shrink compared to the previous case. 

\begin{figure}[tb!]
    \centering
    \includegraphics[width=\textwidth]{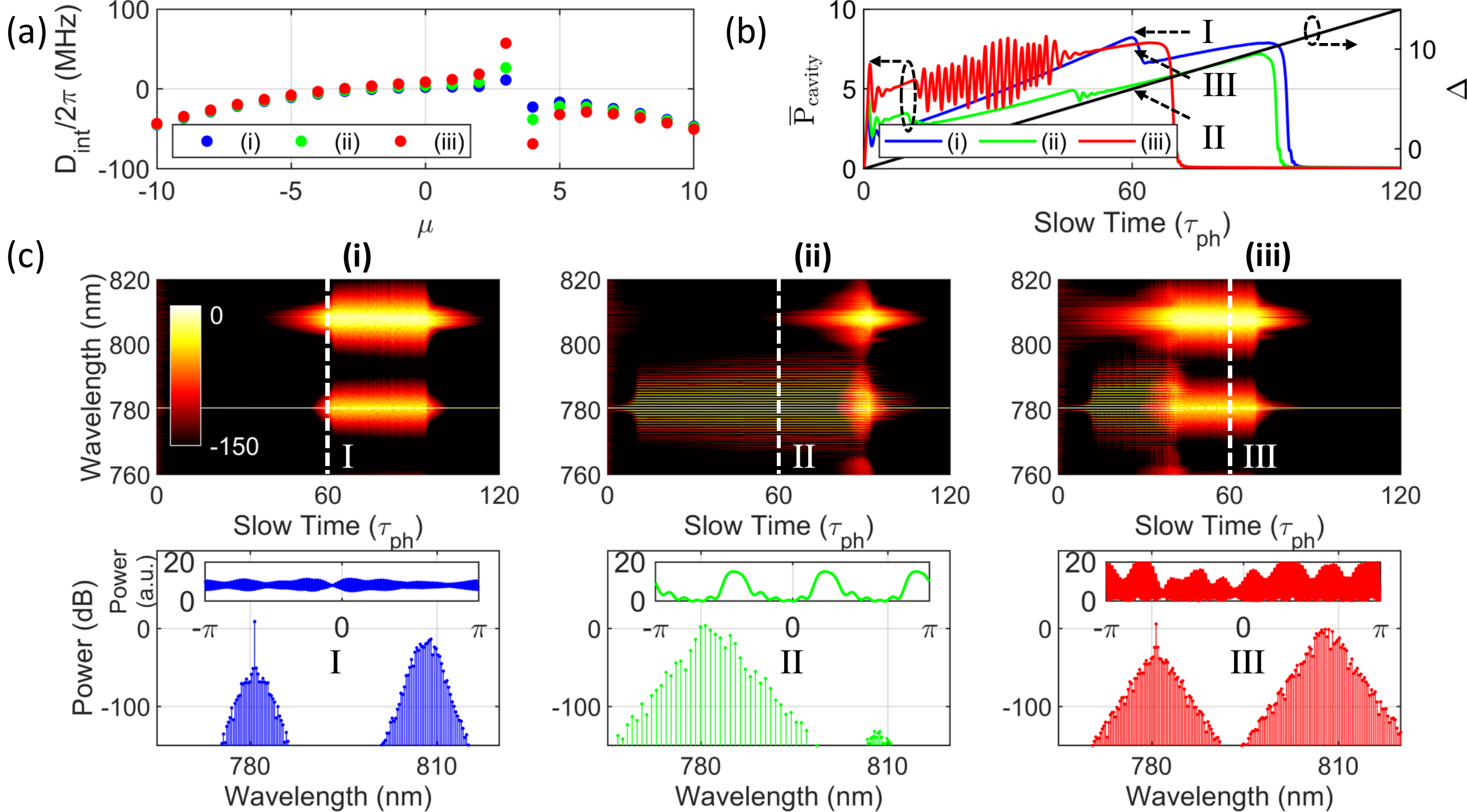}
    \caption{Exication dynamics of dark soliton and SRS at different AMX conditions. (a)~The integrated dispersion with an AMX based on equation~(\ref{eq:7}). Parameters for the AMX are $a=4,8,16$ for case (i), (ii), (iii), respectively, and $b=3$ for  all cases. (b)~The averaged intracavity power for cases (i)~(blue), (ii)~(green), (iii)~(red) and detuning (black) as a function of time. (c)~Spectral evolution of the intracavity fields (top), a representative spectrum (bottom), and temporal waveform (inset) for each case. Different AMX conditions yield different accessible states~(either dark soliton or SRS dominant state).}
    \label{fig:3}
\end{figure}

The normalized coefficient~($a$) for the integrated dispersion, $D_\mathrm{int}(\mu,a,b)$, is chosen to be 4, 8, and 16 for cases~($\mathrm{\textbf{i}}$), ($\mathrm{\textbf{ii}}$), and ($\mathrm{\textbf{iii}}$) at the fixed location $b=3$, respectively, as shown in Figure~\ref{fig:3}(a). The normalized detuning is increased from -2 to 14. The averaged intracavity power is shown in Figure~\ref{fig:3}(b) for each case. The spectral evolution profiles are shown in Figure~\ref{fig:3}(c). In case the parametric gain is high and wide, FWM may be effectively excited even though Raman gain is present and solitons can be generated ~(case~($\mathrm{\textbf{ii}}$) in Figure~\ref{fig:3}(c)). In other words, because nonlinear frequency conversion~(here, FWM) consumes the intracavity power, 
it is required to further increase the detuning to reach the threshold intracavity power for the SRS,
yielding a large existence range for the soliton. The spectrum and temporal profile are also shown which confirms three pulses in the cavity. However as the parametric gain gets lower and narrower, FWM may be dominant over SRS for relatively limited conditions or cannot be excited~(cases~($\mathrm{\textbf{i}}$) and ($\mathrm{\textbf{iii}}$) in Figure~\ref{fig:3}(c)). Then, SRS may be excited easily and a complex interaction between them can occur which often leads to a chaotic temporal profile with Raman oscillation. 

It is critical to introduce an appropriate AMX to access the dark soliton regime. The location can be chosen simply by changing the wavelength of the pump source. The amplitude of the frequency shift may be tuned by indirectly an auxiliary resonator with a microheater~\cite{xue_normal-dispersion_2015,kim_turn-key_2019,helgason_dissipative_2021}, or directly controlling temperature of a cavity with a high thermo-optic coefficient~\cite{shu_sub-milliwatt_2021} or through coupling an auxiliary light into a resonance~\cite{li_thermal_2022}. It is worth mentioning that an oscillatory behavior in the intracavity power for the case~($\mathrm{\textbf{iii}}$) is shown in red in figure~\ref{fig:3}(b) which may be interpreted as a dark breather~\cite{godey_stability_2014, bao_observation_2018}. 

\subsection{Influence of pump power on dynamics of dark soliton generation} 

As the Raman gain does not depend on the AMX as discussed in previous sections, there is a threshold power for SRS regardless of the AMX condition. However the threshold power for FWM is contingent on the AMX effects. In fact, the threshold power for the parametric oscillation can be lower than the threshold power for the Raman oscillation. In this case, the dark soliton regime can be accessed by pumping the cavity with the power between the two threshold powers. As seen in Figure~\ref{fig:1}(a) the parametric gain is in the shape of a semi-ellipse and has a certain existence range as a function of the power, while the Raman gain increases linearly with the intracavity power. This implies that although the Raman effect may be dominant at high pump power, we may find a FWM dominant region at relatively low pump power.

\begin{figure}[tb!]
    \centering
    \includegraphics[width=\textwidth]{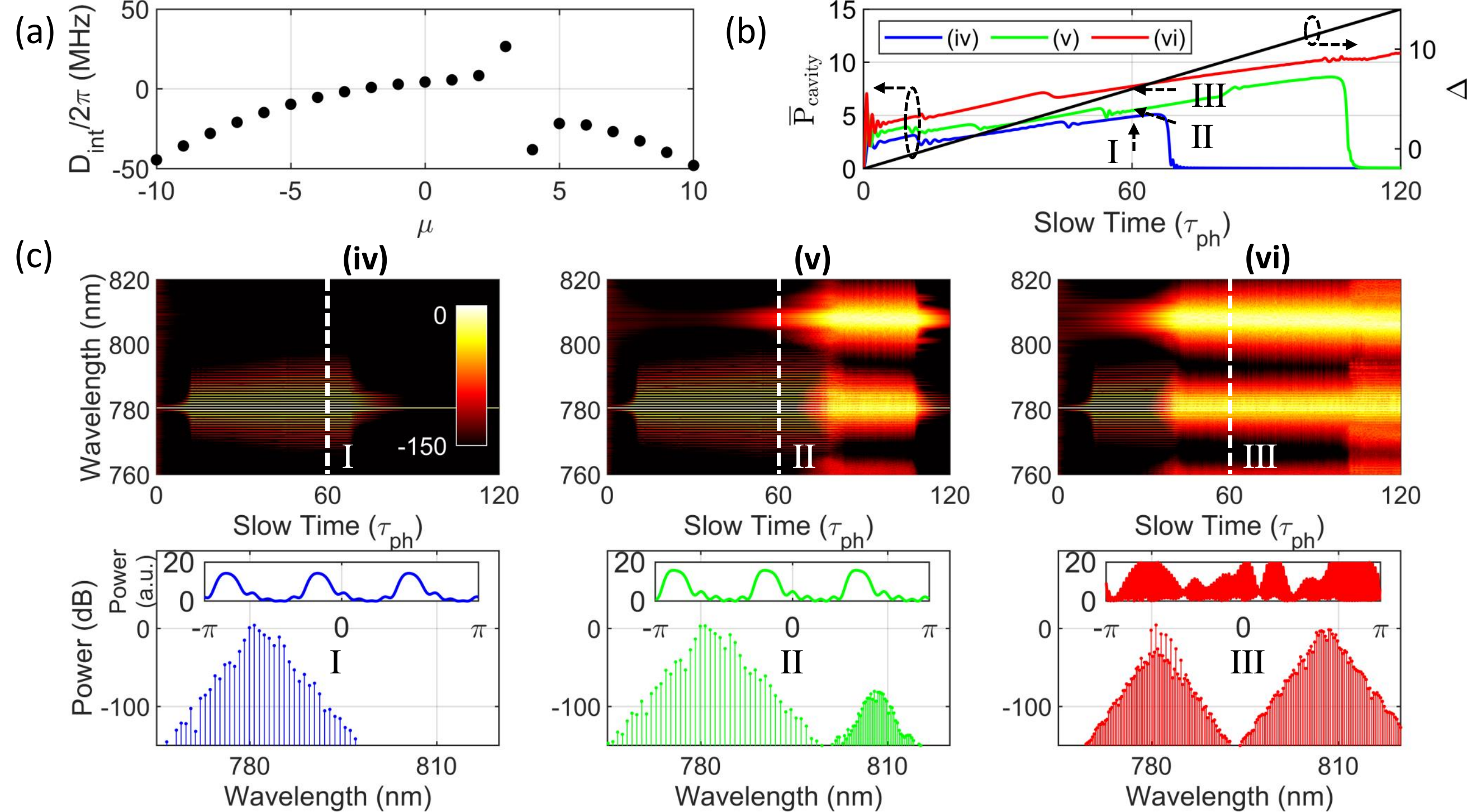}
    \caption{Excitation dynamics of dark soliton and SRS at different pump powers. (a)~The integrated dispersion with an AMX based on equation~\ref{eq:7}, $D_\mathrm{int}(\mu,8,3)$, for all cases. (b)~The averaged intracavity power for cases (iv)~(blue), (v)~(green), (vi)~(red) and detuning (black) as a function of time. (c)~Spectral evolution of the intracavity fields (top), a representative spectrum (bottom), and temporal waveform (inset) for each case. The dark soliton exists for a shorter detuning range at a higher pump power.}
    \label{fig:4}
\end{figure}

The parameters $a$ and $b$ for the dispersion profile are fixed to 8 and 3, respectively, to focus on the effects of power. We choose three different normalized pump powers~($S$ = 3, 5, and 7)  for simulations as shown for cases~($\mathrm{\textbf{iv}}$), ($\mathrm{\textbf{v}}$), and ($\mathrm{\textbf{vi}}$) in figure~\ref{fig:4}(c) while keeping the other parameters the same, respectively. As shown in Figure~\ref{fig:1}(e), we examine three cases: a FWM dominant case~($\mathrm{\textbf{iv}}$), an intermediate case~($\mathrm{\textbf{v}}$), and a SRS dominant case~($\mathrm{\textbf{vi}}$). Provided that the pump power is below the SRS threshold but above the FWM threshold power, a dark soliton state can be accessed for a range of detuning conditions with no observation of the SRS effect, as shown for case~($\mathrm{\textbf{iv}}$) in Figure~\ref{fig:4}(c). Once the pump power is above the SRS threshold power, SRS can be excited as the intracavity power increases and eventually the system will go to a chaotic state, as shown in Figure~\ref{fig:4}(c) for case~($\mathrm{\textbf{v}}$). As the pump power gets higher, SRS will be excited at low detuning values and the soliton state cannot be accessed, as shown in Figure~\ref{fig:4}(c) for case~($\mathrm{\textbf{vi}}$).  Note that a Raman comb can be observed around $810~\mathrm{nm}$, but no evidence of pulse-like pattern can be found in this study~\cite{cherenkov_raman-kerr_2017}. 

In experimental situations, choosing an appropriate pump power and detuning is highly desired to effectively suppress SRS and generate a dark soliton only. However, depending on the AMX condition, it may never be possible to initiate the FWM effect via the mode-interaction-aided parametric gain (Figure~\ref{fig:3}). In case the modal coupling condition cannot be controlled, increasing the threshold power for SRS or decreasing the threshold power for FWM may lead FWM to be a dominant process over SRS, which is demonstrated via a chemical method~\cite{shen_low-threshold_2018} or by adjusting the coupling condition between the cavity and the waveguide~\cite{kippenberg_kerr-nonlinearity_2004,min_controlled_2005}. While no dark breather is observed in this case, a large AMX strength may excite the breather state~\cite{bao_observation_2018}.

\subsection{Stability charts} 
Because there are two important parameters, the detuning and the pump power, different excited states of the system may be plotted in a two-dimensional parameter plane at different modal coupling conditions. As we are only interested in soliton states, the existence region of the soliton can be marked in the plane, which is called a stability chart. This can give us insights on the dark soliton existence and experimental guidelines. To analyze the stability of soliton states for a certain detuning and pump power, the intracavity field is propagated using the LLE. For a fixed pump power, the detuning is increased from -1 to 12 in a discrete step of 0.1~\cite{jaramillo-villegas_deterministic_2015}. In each step, we allowed enough time (here, $30~\tau_\mathrm{ph}$) for the field to pass transitory behavior from a sudden detuning increase and converge to a solution. We recorded the evolution of the field for another period of time (here, $20~\tau_\mathrm{ph}$). Then, this process is repeated for a different pump power. A soliton state is found if the intracavity field of the frequency comb remains constant for the recorded period of time. If Raman lasing occurs, the state is labeled as an SRS state. Although a dark soliton state can be present in the presence of Raman lasing, we exclude this scenario for simplicity as it quickly collapses into a chaotic state as shown in Figure~\ref{fig:3}(c). 

\begin{figure}[tb!]
    \centering
    \includegraphics[width=\textwidth]{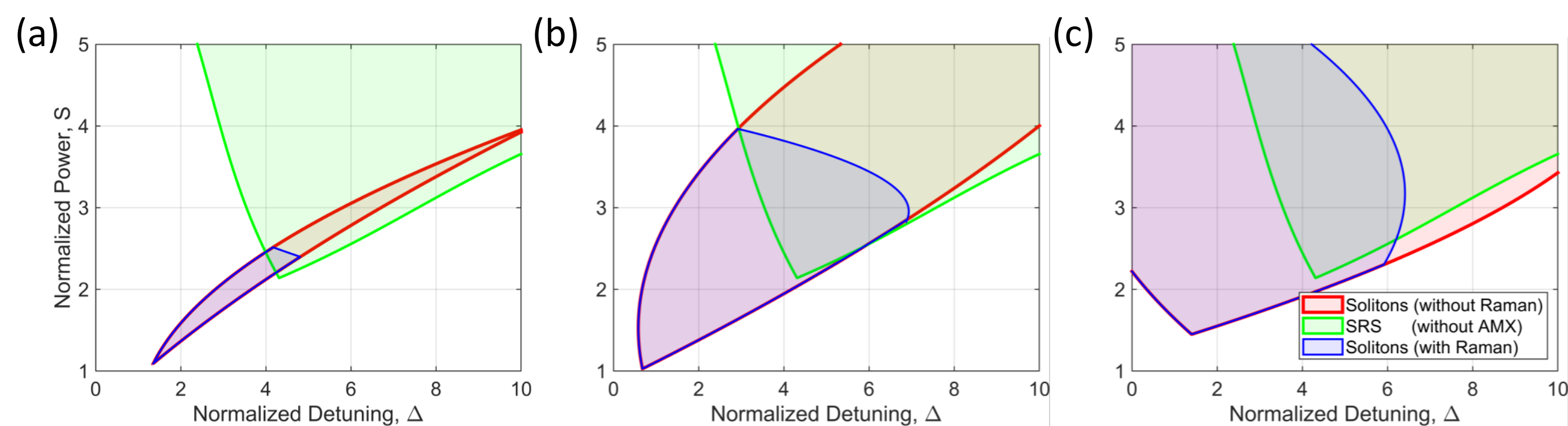}
    \caption{Simulated stability chart for different additional frequency shift values. (a~-~c)~The blue region represents the existence range of dark soliton states in the presence of the Raman interaction~($f_\mathrm{R}=0$.18) for various additional frequency shift values of (a)~$a=1$, (b)~$a=2$, (c)~$a=4$. The red region where dark soliton states exist in the absence of the Raman interaction~($f_\mathrm{R}=0$) and the green region where SRS  is excited in the absence of the additional frequency shift are shown for comparison. The existence range for dark soliton states with the Raman interaction~(blue) is narrower than the one without the Raman interaction~(red). The blue region increases along with the additional frequency shift, but decrease after its maximum value~(not shown here).}
    \label{fig:5}
\end{figure}

Figure~\ref{fig:5} shows regions of stable soliton states~(blue and red) and SRS states~(green) at different AMX conditions. The AMX location~($b$) is fixed to 1 and the strength~($a$) is set to 1, 2, and 4 for the stability analysis which are shown in figures~\ref{fig:5}(a), (b), and (c), respectively. The dark soliton existence region without the Raman interaction is red-colored. The green region represents the presence of the Raman lasing without the modal coupling. Finally, the blue region describes the existence of dark soliton states with the Raman interaction for an AMX condition. When the AMX is relatively small~($a=1$), the existence range of soliton states is narrow without the Raman effect~(red) and even gets narrower with the effect in normal dispersion regime~(blue) as it is analyzed in elsewhere~\cite{godey_stability_2014,parra-rivas_dark_2016,parra-rivas_origin_2016,parra-rivas_influence_2021}. Interestingly, when the AMX strength is relatively large~($a=2,4$), the stable region becomes bigger and even compatible with the region in anomalous dispersion regime. As shown in Figure~\ref{fig:1}, a relatively large AMX strength can introduce a bigger parametric gain region which yields a wider region of soliton states. Thus, it is desired to introduce a relatively significant AMX-induced frequency shift to expand the stable region of the soliton states. Then, we can choose an appropriate pump power and detuning based on the stability chart. 

In practice, unfortunately, it is not trivial to introduce a large modal coupling and control it in a single microresonator. While it is demonstrated that the AMX effects can be controlled by employing a main and an auxiliary microresonators (or coupled microresonators) in an integrated platform ~\cite{miller_tunable_2015,helgason_dissipative_2021,zhu_modulation_2022}, the coupled microresonators may not be easily employed in WGM-type resonators due to difficulties in fabrication. It might be desired that directly controlling the temperature of the cavity without the additional cavity through an auxiliary light~\cite{li_thermal_2022}.  
\section{Conclusion} \label{sec:5}

The interaction of FWM based parametric process and SRS process is investigated in a normal-dispersion microresonator at NIR wavelengths. The phase matching requirement for the parametric process is satisfied by an additional phase shift due to the additional frequency shift caused by an intermodal interaction. It is shown that parametric gain can be created by the additional phase shift. Raman gain is inversely proportional to  wavelength, which makes the competition between the parametric and Raman oscillation more complex at shorter wavelengths. It is found that a significant phase shift can expand and increase the parametric gain envelope which may be beneficial for accessing the dark soliton state. The dynamics of dark soliton generation is analyzed by solving the LLE numerically at various pump powers and modal coupling conditions. An additional frequency shift can excite a dark soliton and extend the existence range, but too big of a frequency shift induces an oscillatory state (or breather) and shrinks the range. The stable region for dark solitons at various pump powers and detuning conditions is summarized in stability charts. 

We believe this work can provide experimental guidelines for AMX based dark soliton generation in systems where the Raman gain is broad and large. Being able to control the AMX condition arbitrarily in a single microresonator (e.g., through temperature control of the cavity) may guarantee dark soliton existence in any microresonator without requiring an auxiliary microresonator, pump modulation, or self-injection-locking.  

\section*{References}
\bibliographystyle{unsrt}
\bibliography{BibTex_100822, JudyBib.bib, JudyBib2.bib}

\end{document}